# Legacy Lending Relationships and Credit Rationing: Evidence from the Paycheck Protection Program

Chunyu Qu[*]

Dec 24, 2025

## Abstract

This article examines how legacy lending relationships shape the allocation of emergency credit under severe information frictions. Using a novel dataset linking Small Business Administration (SBA) loan records with Dun & Bradstreet microdata for over 26 million U.S. firms, I investigate whether prior participation in the SBA 7(a) program acted as a gateway to the Paycheck Protection Program (PPP). Employing entropy balancing to construct a strictly comparable counterfactual group, I document a distinct dynamic evolution in credit rationing. In the program's initial "panic phase" (April 2020), banks relied heavily on legacy ties as a screening technology: firms with prior 7(a) relationships were approximately 29 percentage points more likely to receive funding than observationally identical non-7(a) firms. By June 2021, however, this "insider advantage" had largely vanished, suggesting that policy adjustments and extended timelines eventually mitigated the initial intermediation frictions. These findings highlight a fundamental trade-off between speed and equity in crisis response: while leveraging existing credit rails accelerates deployment, it systematically excludes informationally opaque borrowers. I discuss policy implications for designing future digital infrastructure to decouple verification from historical lending relationships.

---

[*] Chunyu Qu holds a Ph.D. in Economics from Fordham University and is currently a Data Scientist with Dun & Bradstreet Inc (D&B), 2900 Esperanza Crossing, Austin, TX 78758 (email; cqu9@fordham.edu).
The author thanks Sophie Mitra for her major feedback throughout the research. He also thanks Philip Shaw, Subha Mani, Andrew Simons, Patricia Gomez-Gonzalez, and Johanna L. Francis from Fordham University for their constructive feedback. Special thanks go to Yunbo Liu, Assistant Professor at CUHK Business School, for his contributions to the modeling. The author further acknowledges Amber Jaycocks, Nalanda Matia, Julia Zhao, Michelle Solomon, and Andrew Bynes at Dun & Bradstreet for their support, and Tuhin Batra, formerly of D&B and now with Atlantic Technological University, Ireland, for their technical insights and helpful discussions. The author acknowledges Qi He of Google LLC for her insightful input on the comparative assessment and selection of financial metrics. Her expertise in designing augmented metric frameworks and in mechanisms optimization provided valuable analytical choices in this study. *All views expressed in this article are solely of the author and do not represent the views of D&B or any affiliated institution.*

# Introduction

The Paycheck Protection Program (PPP) represented the largest fiscal intervention in U.S. history, channeling over $800 billion in forgivable loans to small businesses. While the program's scale was unprecedented, its implementation faced a classic economic dilemma: how to rapidly allocate liquidity in a crisis when lenders face severe information frictions regarding borrower viability. In the absence of perfect information, banks often rely on "relationship lending" as a screening technology. This article examines how legacy lending relationships, specifically prior participation in the Small Business Administration's (SBA) 7(a) program, acted as a decisive, yet unequal, gateway to emergency credit.

Existing literature on the PPP has largely focused on the supply side, documenting how bank branch networks (Granja et al. 2022), lender technology (Howell et al. 2024), and banking relationships (Li and Strahan 2021) shaped aggregate allocation. While these studies highlight geographic and institutional disparities, they often view the problem through the lens of the lender. Much less is known about the firm-level selection mechanisms: specifically, how a firm's pre-crisis "institutional resume", its history of government-backed borrowing and verified creditworthiness, determined its survival in the race for funds.

This article fills this gap by constructing a novel, near-universe dataset linking SBA records to microdata from Dun & Bradstreet (D&B). I merge PPP loan records with historical SBA 7(a) data and D&B credit files for over 26 million U.S. firms. This granularity allows for a precise examination of how legacy government credit relationships interact with firm fundamentals, such as commercial credit and financial stress, to determine credit access.

Distinguishing this analysis from prior work, I employ entropy balancing, a rigorous reweighting method that improves upon standard propensity score matching, to estimate the causal effect of prior 7(a) participation. This approach ensures that the estimated advantages of legacy borrowers are not merely artifacts of their size or observable quality, but reflect the specific value of established institutional ties.

The empirical results reveal a striking dynamic evolution in credit allocation, highlighting a "learning curve" in crisis response. First, in the chaotic early phase (April 2020), PPP access was heavily skewed toward "insiders." Firms with prior 7(a) relationships were approximately 29 percentage points more likely to receive funding than observationally identical non-7(a) firms. During this period, banks effectively used 7(a) history and high credit scores as a proxy for verification speed, rationing credit away from opaque, albeit viable, borrowers.

Second, this screening mechanism relaxed significantly over time. By June 2021, as the program matured and rules were adjusted, the predictive power of legacy ties and credit scores largely vanished. Late-stage recipients resembled the general population of small businesses, suggesting that policy adjustments eventually mitigated the initial information frictions.

This study makes two primary contributions to the literature on small business finance and crisis management. First, it quantifies the "program-layering premium." I show that participation in peacetime guarantee programs (like 7(a)) creates an embedded advantage for accessing future emergency aid, a mechanism that race-neutral policies may inadvertently reinforce. Second, it provides a firm-level dynamic assessment of the trade-off between speed and equity. While leveraging legacy relationships expedited initial disbursements, it created significant access inequality that was only corrected in later phases. These findings suggest that future emergency lending infrastructures must move beyond reliance on legacy banking lists toward more inclusive, pre-qualified digital verification systems.

## Section II. Literature Review

A rapidly growing literature examines how PPP funds were allocated and what effects they had on firms and local economies. Autor et al. (2022b) provide an overview of the program's design and aggregate distribution, showing that a large share of subsidies accrued to business owners and high-income households rather than directly preserving jobs. Hubbard and Strain (2021), Dalton (2023) and Autor et al. (2022a) use administrative and payroll microdata to place PPP in the broader context of pandemic business support, finding modest employment effects relative to the program's fiscal cost.

One strand emphasizes bank supply, relationships, and information frictions. Humphries, Neilson, and Ulyssea (2020) use survey data to document that many small firms, especially those without strong banking ties, faced confusion about eligibility, difficulty finding participating lenders, and uncertainty around forgiveness, leading to substantial take-up frictions. Li and Strahan (2021) show that PPP banks with more pre-existing small-business relationships and greater engagement in commitment lending supplied more PPP loans. Granja et al. (2022) show that funds flowed toward regions with denser branch networks and better-capitalized banks. Joaquim and Wang (2022) show that recipiency of a loan substantially improved subsequent financial condition.

A second strand studies distributional disparities and the role of lender technology. Atkins, Cook, and Seamans (2022) show that Black-owned firms received smaller PPP loans than otherwise similar firms. Chernenko and Scharfstein (2024) and Chernenko et al. (2023) find that both lower application rates and differential approval contribute to gaps, with administrative burdens and confusion playing important roles. Howell et al. (2024) show that fintech and highly automated lenders were more likely to serve minority-owned firms and to extend smaller loans.

These literatures provide a rich picture of where PPP dollars went and how lender behavior affected targeting, however, three gaps remain. First, we know relatively little about how legacy program participation, in particular prior SBA 7(a) borrowing, shaped access to PPP at the firm level, even though 7(a) relationships are a natural policy lever and a core feature of the U.S. small-business credit infrastructure. Second, existing work rarely traces how the mapping from firm characteristics to PPP access changed over the life of the program: most analyses focus on early waves rather than the whole PPP window. Third, few studies combine near-universe firm-level data with modern reweighting methods to approximate experimental comparisons between firms with and without prior program ties.

This article addresses these gaps by: (i) placing prior SBA 7(a) participation at the center of the analysis as a predetermined, institutionally meaningful state variable; (ii) documenting how the joint roles of 7(a) ties, credit quality, financial stress, and firm size in shaping PPP access evolved between April 2020 and June 2021; and (iii) applying entropy-balancing ATT estimators to a merged D&B–SBA–PPP dataset covering virtually the entire U.S. employer firm population. In doing so, it complements existing bank- and geography-focused studies with a firm-level account of how legacy government programs and private credit hierarchies interacted to determine access to crisis-era support.

## Section III. Data and Variables

This study combines firm-level data from Dun & Bradstreet (D&B) with administrative records from the PPP and SBA's 7(a) lending program. Detailed information on matching procedures, variable construction, and additional descriptive statistics is provided in Appendix A. To measure prior SBA borrowing experience, I use SBA 7(a) loan data from 2010–2019 and match these records to D&B firms. This yields a binary indicator of whether a firm had obtained at least one 7(a) loan prior to the pandemic.

### A. Outcome variables

I consider two main outcome variables:

PPP loan recipiency. A binary indicator equal to one if a firm is matched to at least one PPP loan by the reference date and zero otherwise.

PPP loan amount. The total PPP loan amount (in U.S. dollars) received by each matched firm.

### B. Treatment and key firm-level covariates

The central treatment variable in the analysis is firms' prior experience with SBA's main small-business lending program:

*SBA 7(a) experience.* An indicator equal to one if a firm received at least one SBA 7(a) loan between 2010 and 2019 and zero otherwise. Because 7(a) loans predate the pandemic and the PPP rollout, this variable captures long-standing relationships with SBA-backed lenders and familiarity with SBA procedures.

In addition, I use several pre-PPP variables from D&B as key covariates to summarize firms' financial condition and organizational scale:

*Commercial Credit Ranking.* A percentile ranking from 1 to 100, where higher values indicate a lower predicted probability of severe delinquency, bankruptcy, or closure over the next 12 months.

*Financial Stress Ranking.* A percentile ranking from 1 to 100, where higher values indicate a lower predicted probability that a firm will seek legal relief or cease operations without repaying creditors within the next 12 months.

*Firm size.* The employment data reported by D&B, updated monthly. I use both the level and its natural logarithm to allow for nonlinear size effects.

*Business age.* The number of years since the firm first entered the D&B database, capturing the maturity and track record of the business.

These variables are predetermined or slow-moving relative to PPP approval decisions and provide a compact representation of firms' creditworthiness, failure risk, and scale. Construction details for the D&B risk rankings are given in Appendix A.2.

### C. Secondary demographic covariates and controls

The analysis also incorporates ownership demographics and additional controls. All regressions include additional controls for organizational form and local economic conditions, such as ownership/lease status, prior business failure flags, foreign or public status, industry codes, and state fixed effects. These variables help absorb broad differences in sectoral composition and geographic exposure to the pandemic. Appendix A.3–A.5 provide formal definitions and additional descriptive statistics for these controls.

## Section IV. Empirical Strategy and Identification

The analysis also incorporates ownership demographics and additional controls. All regressions include additional controls such as ownership/lease status, prior business failure flags, foreign or public status, industry codes, and state fixed effects. These variables help absorb broad differences in sectoral composition and geographic exposure to the pandemic. Appendix A.3–A.5 provide formal definitions and additional descriptive statistics for these controls.

## A. Baseline models

**(1) PPP recipiency**

I describe the econometric model used to investigate the association between PPP recipiency and selected characteristics. The baseline linear probability model is

$$Y_{ijs} = \beta'Z_i + \gamma'X_i + \alpha_j + \rho_s + \varepsilon_{ijs}. \tag{1a}$$

where $Z_i$ can be $7a_i$ (indicator for having at least one SBA 7(a) loan between 2010–2019), commercial credit and financial-stress rankings (1–100, higher values = lower risk), $\ln(firmSize)$ (log number of employees), and $Age_i$ (firm age). Let $X_i$ collect additional controls: ownership status, business-failure history, foreign and public status, county population, log sales, subsidiary indicator, central-city and rural/suburban region dummies, and other structural characteristics. All specifications include state fixed effects $\rho_s$ and broad industry fixed effects $\alpha_j$.

Equation (1a) is estimated by OLS with standard errors clustered at the county level. Coefficients on $Z_i$ are interpreted as percentage-point changes in PPP recipiency associated with a unit change in the corresponding characteristic, conditional on the full set of controls and fixed effects. For expositional clarity, I also report a set of specifications where equation (1a) is re-written as

$$Y_{ijs} = \beta D_i + \sum_k \gamma_k X_{ik} + \alpha_j + \rho_s + \varepsilon_{ijs}, \tag{1b}$$

and the "highlighted" regressor $D_i$ is rotated across the key interpretive variables, while the remaining firm characteristics stay in $X_{ik}$. Columns based on (1b) make it easier to compare the marginal association of each characteristic with PPP recipiency, but the joint specification (1a) is treated as the preferred baseline.

**(2) PPP loan amounts**

To study the intensive margin among PPP recipients, let LoanAmt$_{ijs}$ denote the total PPP loan amount received by firm $i$. The baseline specification for loan amounts is the following OLS model in levels:

$$\text{LoanAmt}_{ijs} = \delta'Z_i + \phi'X_i + \alpha_j + \rho_s + u_{ijs}. \tag{2a}$$

δ is a vector of coefficients on the core characteristics, ϕ collects coefficients on the additional controls, the other terms keep still. Similarly, (1b) turns to be

$$\text{LoanAmt}_{ijs} = \theta D_i + \sum_k \kappa_k X_{ik} + \alpha_j + \rho_s + v_{ijs}, \tag{2b}$$

with all the terms the same as (1b). (2b) facilitate direct comparison of the marginal association between each firm characteristic and the dollar level of PPP support, while (2a) remains the preferred model.

## B. Identification Strategy

To move closer to a causal effect of 7(a) experience on receiving PPP loans, I adopt a treatment–effects framework that compares firms with and without 7(a) experience that are otherwise similar in observed pre-PPP characteristics. The goal is to estimate the average effect of 7(a) relationships on PPP access for firms that actually had such relationships.

Let $T_i = \mathbf{1}\{loan7a_i = 1\}$ indicate whether firm $i$ had at least one SBA 7(a) loan approved between 2010 and 2019, and let $Y_i = 1\{PPP_i = 1\}$ be an indicator for receiving at least one PPP loan by the reference date. The causal estimand of interest is the average treatment effect on the treated (ATT),

$$ATT = E[Y_i(1) - Y_i(0) \mid T_i = 1], \tag{3a}$$

where $Y_i(1)$ and $Y_i(0)$ denote firm $i$'s potential PPP recipiency with and without prior 7(a) borrowing. Identification relies on two standard assumptions: (i) selection on observables, $\{Y_i(0), Y_i(1)\} \perp T_i \mid X_i$, where $X_i$ collects f covariates, and (ii) overlap in the propensity to obtain 7(a) loans, $0 < \Pr(T_i = 1 \mid X_i) < 1$ for firms in the support of $X_i$.

The covariate vector $X_i$ includes the core firm characteristics used in the baseline regressions, commercial credit ranking, financial-stress ranking, log firm size, firm age, and log sales, together with indicators for premise ownership, prior business-failure history, foreign and public status, metropolitan and rural/suburban location, and industry and state fixed effects. All covariates are measured prior to the start of the PPP program.

For the main causal estimates, I use entropy balancing to construct a synthetic control group of non-7(a) firms whose covariate distribution matches that of 7(a) firms. Intuitively, rather than discarding most controls and keeping only a few nearest neighbors, entropy balancing retains the full sample of non-7(a) firms but assigns them continuous weights so that their weighted covariate means coincide with those of the treated group.

Formally, let $w_i$ denote the weight assigned to firm $i$. Treated firms ($T_i = 1$) receive unit weight, $w_i = 1$. For control firms ($T_i = 0$), entropy balancing chooses non-negative weights $w_i$ that minimize the Kullback–Leibler divergence from uniform weights subject to linear balancing constraints on $X_i$:

$$\min_{w_i : T_i = 0} \sum_{i:T_i=0} w_i \log\left(\frac{w_i}{\bar{w}}\right) \tag{3b}$$

subject to

$$\sum_{i:T_i=0} w_i X_i = \sum_{i:T_i=1} X_i, \tag{3c}$$

and $\sum_{i:T_i=0} w_i = N_T$, where $\bar{w}$ is a normalizing constant and $N_T$ is the number of treated firms. In practice, I impose balance on the first moments of all covariates listed in Table 4 and verify robustness to also balancing selected second moments. Panel A of Table 4 shows substantial raw imbalance between 7(a) and non-7(a) firms in credit rankings, financial-stress rankings, size, age, and ownership structure. Panel B demonstrates that the entropy-balancing weights reduce all standardized differences to below 5 in absolute value, indicating excellent covariate balance. Given the entropy-balancing weights, the ATT on PPP recipiency is estimated as a weighted difference in mean outcomes:

$$\widehat{ATT}^{EB} = \frac{1}{N_T}\sum_{i:T_i=1} Y_i - \frac{1}{N_T}\sum_{i:T_i=0} w_i Y_i. \tag{3d}$$

using the entropy-balancing weights for control firms and unit weights for treated firms, and interpret $\hat{\tau}$ as a regression-adjusted ATT. Standard errors are computed using robust variance estimators with clustering at the county level. Table 5 reports the resulting ATT estimates for PPP recipiency; they remain large and highly significant and are only moderately smaller than the corresponding baseline LPM coefficients, suggesting that a substantial 7(a) advantage persists even after aligning the observable characteristics of 7(a) and non-7(a) firms.

## C. Robustness checks

As a robustness check for model (1) and (2), I estimate probit and logit models for PPP recipiency and report average marginal effects. The signs and magnitudes closely mirror the LPM results; see Appendix B.

As a robustness check for causal inference model (3), I also implement nearest-neighbor propensity-score matching. I first estimate the propensity to obtain a 7(a) loan, $\Pr(T_i = 1 \mid X_i)$, using a logit model with the same covariates $X_i$ as in the entropy-balancing design. Each treated firm is then matched to the nearest untreated firm in terms of the estimated propensity score, within a common-support region. The ATT is computed as the average difference in PPP recipiency between treated firms and their matched controls. The matching-based ATT estimates, reported in Appendix D, are very close in magnitude to the entropy-balancing estimates, reinforcing the conclusion that prior 7(a) relationships had a sizeable causal impact on PPP access for otherwise similar firms.

# Section V. Results

## A. Descriptive Statistics

Tables 1 and B1 describe firm characteristics in April 2020 and June 2021, separately for PPP recipients and non-recipients. In April 2020, 26.7 million firms are observed, of which about 22.9 million received a first-draw PPP loan. Recipients are much more likely to have prior 7(a) borrowing, display higher commercial-credit rankings and lower predicted financial stress, and are larger and older on average than non-recipients. By June 2021, the sample expands to 29.8 million firms, but only 1.6 million obtain a first-draw loan in that later window, and the gaps in credit quality, stress rankings, firm size, and age between recipients and non-recipients shrink sharply. These patterns suggest that early PPP lending was concentrated among better-established, lower-risk firms, while late-stage recipients more closely resemble the broader small-business population. Moreover, PPP recipiency exhibits an inverted-U relationship with firm size: mid-sized firms (101–250 employees) have the highest probability of receiving a loan, whereas very small and very large firms are much less likely to obtain PPP funds (Figure 1).

The descriptive statistics show that early PPP funds flowed disproportionately to larger, older, lower-risk firms with prior SBA relationships, whereas by mid-2021 the marginal PPP recipient was closer to the typical small business in the population, both in financial condition and in observables. The regression and matching analyses that follow quantify these patterns and examine how much of the early advantage associated with 7(a) relationships and stronger credit profiles persists after conditioning on observables.

Finally, the monthly timing patterns reinforce the layering interpretation and highlight how quickly 7(a) firms reacted. Table C2 shows that about 50% of all PPP recipients with prior 7(a) borrowing obtained their first PPP loan in April 2020, compared with roughly 32% of all PPP recipients. Over the full first year, around 69% of 7(a)–PPP firms entered the program in 2020 versus 55% of all PPP recipients. By contrast, in the 2021 rounds the monthly share of new PPP loans going to 7(a) firms is consistently below their share in the overall PPP pool. This front-loaded pattern indicates that legacy SBA participants not only had higher approval chances, but also moved into the program much earlier, when capacity constraints and processing bottlenecks were most severe.

## B. Baseline Regressions

Table 2 presents the estimates for the program's initial phase (April 2020), a period characterized by extreme liquidity demand and administrative bottlenecks. The results indicate that credit allocation was heavily governed by pre-existing institutional ties and verifiable creditworthiness. As shown in column (6), having a prior SBA 7(a) relationship raised the probability of receiving a PPP loan by approximately 31.6 percentage points, conditional on all controls. This magnitude suggests that during the "panic phase" of the crisis, banks effectively rationed credit by prioritizing "known" borrowers to minimize processing time and information costs. Similarly, commercial credit ranking and firm size served as strong positive predictors, reinforcing the view that early allocation followed a "flight to quality" pattern, where informationally opaque or riskier firms were systematically crowded out.

By June 2021, the allocation landscape had shifted fundamentally. Table 3 reveals that the screening mechanisms dominant in the early rounds had largely dissipated. The coefficient on 7(a) experience drops to near zero (and turns slightly negative), implying that the initial advantage of legacy borrowers did not persist into the program's maturity. Furthermore, the marginal effects of credit rankings and financial stress scores shrank by an order of magnitude compared to April 2020. This convergence indicates that as the liquidity crisis eased and policy rules were relaxed to encourage broader participation, the "information rent" associated with legacy banking relationships was competed away. The distinct inverted-U relationship between firm size and recipiency, visualized in Figure 2(a), further confirms that while mid-sized firms were served early, the program eventually saturated the eligible population, reaching the smallest micro-firms and those with weaker credit profiles in later waves.

Comparing the coefficients across the two periods highlights a critical time-varying trade-off in emergency lending. The sharp decline in the predictive power of 7(a) ties and credit scores (from Table 2 to Table 3) suggests a rapid relaxation of information frictions. In April 2020, legacy relationships acted as a necessary screening technology, a "fast track" that facilitated speed but generated inequality. By June 2021, the disappearance of these effects signals that the combination of fintech entry, policy recalibration, and extended deadlines successfully mitigated the initial rationing. This temporal heterogeneity challenges static evaluations of the PPP; it suggests that "misallocation" was largely a feature of the program's initial design for speed, which was subsequently corrected as the mechanism of delivery evolved from relationship-based triage to broad-based distribution.

Credit quality and failure risk also matter, but on a smaller scale. A one-point increase in the commercial-credit ranking (on its 1–100 scale) is associated with roughly a 0.10 percentage-point increase in the probability of receiving a loan, so moving a firm one decile up the national credit distribution raises PPP access by about 1 percentage point. The financial-stress ranking has a similar positive association. Firm size is another key predictor: in column (6) the coefficient on log firm size implies that a one-log-point increase in firm size (approximately a 170% firm size increase) raises the probability of PPP recipiency by about 4–5 percentage points. Firm age plays a minor role once size and risk are controlled for.

The "rotated" specifications in columns (2)–(5) confirm that each characteristic is individually powerful: when entered one at a time, higher credit ranking, lower financial stress, and larger firms are all significantly more likely to obtain PPP loans. Importantly, the magnitudes in column (6) are close to those in columns (1)–(5), suggesting that multicollinearity is not driving the results and that the large 7(a) effect reflects an incremental advantage even after conditioning on observable risk and size. Overall, these regressions reinforce the descriptive patterns: in the program's first month, PPP loans flowed disproportionately to larger, older, lower-risk firms with existing SBA relationships.

By June 2021, these patterns change dramatically. In Table 3, column (6), the coefficient on 7(a) experience is now small and slightly negative (around –0.8 percentage points), indicating that among firms still applying for first-draw loans in mid-2021, having prior 7(a) borrowing no longer confers an advantage and may even be mildly associated with lower recipiency. The coefficients on credit and financial-stress rankings remain positive but are an order of magnitude smaller than in April 2020: a 10-point improvement in either ranking now corresponds to less than a one-percentage-point increase in PPP access. The effect of firm size nearly vanishes and turns slightly negative, consistent with the program having largely saturated larger firms earlier.

Taken together, the April 2020 and June 2021 regressions show a clear evolution. Initially, PPP allocation was heavily shaped by pre-existing SBA relationships, credit quality, and firm size, with 7(a) borrowers enjoying a very large edge. Over time, as program rules evolved and remaining applicants became more marginal, these advantages eroded: the predictive power of 7(a) status and risk rankings weakened sharply, and PPP recipiency among new entrants in 2021 looked much closer to that of the average firm in the economy. The next subsection uses propensity-score and entropy-balancing methods to test whether the early 7(a) advantage survives when comparing treated and untreated firms with very similar observable characteristics.

## C. Intensive margin - Loan amounts

Tables B4 and B5 report OLS estimates of equation (2a) for the dollar amount of PPP loans among recipient firms. In April 2020, prior 7(a) participation is associated not only with a higher probability of receiving a PPP loan but also with substantially larger loans conditional on recipiency. In the full April specification, firms with 7(a) experience receive on average about $25,000 more in PPP funding than comparable non-7(a) firms, roughly a 20 percent increase relative to the mean first-draw loan. Higher commercial-credit rankings, lower financial stress, and larger firm size are also associated with larger loan amounts, in line with the program's payroll-based formula.

By June 2021, this intensive-margin advantage has essentially disappeared. In the June specification, the 7(a) coefficient is close to zero and statistically insignificant, indicating no meaningful difference in loan size between 7(a) and non-7(a) firms among late-stage recipients. The coefficients on credit rankings, financial stress, and firm size remain positive but are smaller in magnitude than in 2020, mirroring the convergence in observable characteristics documented in the descriptive statistics.

## D. Causal impact of 7(a) relationships on PPP access

Table 4 documents how firms with prior 7(a) borrowing differ from those without such relationships before any reweighting. Panel A shows substantial raw imbalance in pre-PPP characteristics. Relative to non-7(a) firms, 7(a) borrowers have much stronger credit rankings and lower predicted financial stress, are larger and older on average, and are more likely to operate in metropolitan locations. The standardized mean differences for credit ranking, financial-stress ranking, log firm size, and firm age all exceed 50 in absolute value, with values around 70 for the two risk scores. These magnitudes indicate strong positive selection of financially healthier, more established firms into the 7(a) program.

Panel B of Table 4 demonstrates that the entropy-balancing weights are highly successful in aligning the observable characteristics of non-7(a) firms with those of 7(a) borrowers. After reweighting, the weighted means for the control group closely track those of the treated group, and all standardized differences fall below 7, with most below 5 in absolute value. In other words, conditional on the rich set of firm-level controls, the weighted non-7(a) sample becomes observationally similar to the 7(a) sample along dimensions of credit quality, financial stress, size, age, ownership, prior failure history, public and foreign status, and location. This provides a transparent check that the entropy-balancing design delivers excellent covariate balance.

Table 5 reports the resulting average treatment effects on the treated (ATT) of prior 7(a) borrowing on the probability of receiving a PPP loan, using the full firm population observed in June 2021. The raw difference in means indicates that 7(a) firms were far more likely to obtain PPP funding: about 60.7 percent of firms with 7(a) experience received a PPP loan, compared with only about 19.6 percent of firms without such experience, yielding an unadjusted gap of roughly 41 percentage points. Once industry and state fixed effects and the full set of firm-level controls are included in the baseline linear probability model, the estimated 7(a) coefficient declines but remains very large at 0.316 (column "OLS"). The causal estimators based on matching and weighting produce slightly smaller but tightly clustered effects: nearest-neighbor propensity-score matching yields an ATT of 0.305, while the entropy-balancing estimator, the preferred specification, implies that prior 7(a) relationships raise the PPP recipiency probability by about 0.292. Relative to the 19.6 percent PPP take-up rate among comparable non-7(a) firms, this implies an increase of roughly 150 percent. The stability of the estimated treatment effect across OLS, matching, and entropy-balancing specifications suggests that, even after accounting for substantial selection on observable credit quality and firm characteristics, pre-existing 7(a) relationships had a large and economically meaningful causal impact on firms' access to PPP funds.

### E. Robustness and Sensitivity Results

As a robustness check, I re-estimate the PPP-recipiency models using probit specifications (Tables B2–B3). The probit coefficients closely mirror the linear-probability results: in April 2020, prior 7(a) experience remains strongly and positively associated with PPP access, while higher commercial-credit and financial-stress rankings and larger firm size also significantly raise the likelihood of receiving a loan. By June 2021, the marginal effect of 7(a) becomes small and negative and the effects of credit, stress, and size all shrink substantially, reproducing the pattern of weakening selection documented in Tables 2–3. Overall, the probit estimates confirm that the main findings are not driven by the choice of a linear-probability model.

Descriptive patterns also reveal pronounced heterogeneity in the underlying 7(a) coverage rate (Appendix Table C1). Only about 0.06 percent of micro firms with 1–5 employees have any 7(a) history in the comprehensive population of US small firms, compared with roughly 1.2 percent of firms with 21–50 workers, before the share declines again for 200+ employee firms. A similar gradient appears in business age, rising from about 0.2 percent among firms aged 1–5 years to 0.8 percent for those older than 15 years, and along pre-PPP credit and sales distributions. Legacy 7(a) relationships were thus already concentrated among larger, older, financially stronger and higher-sales businesses before the pandemic, which helps explain why these firms enjoyed disproportionate PPP access once the program opened.

## Section VI. Discussion and Implications

### A. Summary of Main Findings and Mechanisms

The evidence suggests three central messages. First, pre-existing SBA 7(a) relationships created a very large and robust advantage in accessing PPP funds. In the raw data for June 2021, roughly 60.7 percent of firms with at least one 7(a) loan had received a PPP loan, compared with only 19.6 percent of firms without such borrowing, implying an unadjusted gap of about 41 percentage points. After conditioning on rich firm-level controls and state and industry fixed effects, the baseline linear probability model still attributes a 31.6-percentage-point higher probability of PPP recipiency to prior 7(a) borrowing. Entropy-balancing weights that align non-7(a) firms with 7(a) borrowers on observables reduce the estimate only modestly, to an ATT of roughly 29 percentage points. This pattern indicates that legacy SBA relationships, rather than observable differences in risk, size, or location alone, had an independent and economically large effect on whether firms accessed PPP funds.

A simple back-of-the-envelope calculation illustrates the policy magnitude of this relationship channel. By the end of PPP (June 2021), there were 187,762 businesses with at least one historical 7(a) loan. Combining this count with the preferred entropy-balancing ATT estimate of 0.292 implies that roughly $0.292 \times 187{,}762 \approx 55{,}000$ additional firms received a first-draw PPP loan because they had a legacy 7(a) relationship rather than observationally similar characteristics alone. Using the average first-draw PPP loan size among June recipients of about USD 20,000, this translates into roughly USD 1.1 billion in incremental, largely forgivable credit effectively earmarked for firms already inside the federal guarantee system. It highlights how legacy program participation can quietly steer tens of thousands of marginal awards and a non-trivial amount of fiscal support toward established borrowers, with obvious implications for the design of future emergency lending schemes.

Second, the strength of selection on 7(a) ties, credit quality, and firm size changed markedly over time. In April 2020, PPP recipiency was heavily tilted toward firms that had prior 7(a) loans, scored high on commercial credit and low on financial stress, and employed more workers. By June 2021, these coefficients shrank dramatically: the 7(a) effect in the baseline regressions becomes small and slightly negative, and the marginal effects of credit, stress, and size are an order of magnitude smaller. Combined with the descriptive statistics, this suggests that the first tranche of PPP funds primarily reached financially stronger, more established firms that were already inside the SBA and banking networks, whereas later rounds extended support to a pool of applicants that more closely resembled the broader small-business population. The causal analysis shows, however, that even in 2021, conditional on observables, firms with 7(a) ties still enjoyed a sizable residual advantage.

Third, the results highlight a speed–targeting–equity trade-off that is central to crisis-era business lending. Early in the pandemic, when avoiding mass layoffs was paramount, channeling funds through entities with existing documentation and relationships may have been the fastest way to move money. Yet doing so entrenched pre-existing hierarchies in credit access and program familiarity. Subsequent reforms and outreach broadened eligibility and diffused information, but they could not fully erase the institutional advantage of firms already integrated into SBA lending programs.

Finally, linking these findings on access to broader program outcomes suggests that the reliance on legacy rails may have generated a "double distortion." In companion work, Qu (2025) finds that the marginal impact of PPP loans on employment retention, financial stress alleviation, and commercial credit improvement was significantly smaller for firms with 7(a) experience compared to their non-7(a) counterparts. This implies that the SBA's infrastructure implicitly prioritized "insiders", who were already financially stronger, over "outsiders" for whom the marginal utility of liquidity was arguably higher. Therefore, the structural inequality documented in this article likely came at the cost of aggregate allocative efficiency, diverting emergency funds away from the firms that could have derived the greatest stabilizing benefit. This reinforces the urgency of decoupling future crisis verification from historical lending relationships, not merely for the sake of equity, but to maximize the economic multiplier of fiscal intervention.

## B. Policy Implications for Crisis-Response Lending

The empirical findings of this article highlight a fundamental structural tension in crisis response: the mismatch between universal liquidity needs and niche transmission channels. While relying on the SBA 7(a) infrastructure allowed for rapid initial deployment, it essentially privatized the allocation decision to commercial banks. As established in the macro-finance literature, financial intermediaries naturally prioritize relationships that minimize their own information costs and credit risks, leading to a "flight to quality" that may not align with the social objective of broad stabilization. To resolve the distortions observed in the PPP's "first-come, first-served" (FCFS) design, future policy must move beyond ad-hoc adjustments toward structural reform in three specific areas.

1. The "Pre-Qualification" Model The reliance on commercial banks to process applications created a bottleneck where firms with pre-existing 7(a) ties were prioritized, while "outsiders" faced administrative queues. To mitigate this intermediation friction, future emergency programs should establish a direct "public option" rail that bypasses bank underwriting for standardized relief.

2. Redesigning Allocation Mechanisms: Alternatives to "First-Come, First-Served" This article shows that the FCFS mechanism acted as a force multiplier for inequality. Because 7(a) borrowers could navigate the application process faster (often with lender assistance), the FCFS rule effectively reserved the initial funding tranche for them, leaving the fund depleted for others. Future programs should abandon strict FCFS for oversubscribed facilities in favor of Algorithmic Tranching or Lottery-Based Processing within the first window (e.g., the first 72 hours). Rather than processing applications linearly, the administrator could enforce predetermined quotas based on firm size, geography, or lack of prior credit history from Day 1. For instance, reserving 20% of the initial tranche specifically for "non-relationship" borrowers would structurally prevent the crowding-out effect observed in April 2020.

3. Investing in Digital Public Infrastructure (DPI) to Reduce Information Asymmetry The core reason banks relied on 7(a) history was the lack of verifiable information on other small businesses. The "water" of liquidity failed to reach the "cracks" of the economy because the plumbing, the information infrastructure, was incomplete. Policymakers should accelerate the development of Open Banking APIs and a standardized Digital Business ID system. Similar to successful implementations in other economies, a unified digital identity that links a firm's tax filing, utility payments, and bank account would allow lenders (including non-bank FinTechs) to verify a firm's existence and scale instantaneously, without relying on manual document review or prior relationship history. By reducing the marginal cost of verifying "unknown" firms, such infrastructure would reduce the lender's incentive to ration credit solely to legacy clients.

4. The reliance on private intermediaries created a principal-agent conflict: while the government's objective was broad coverage, the lenders' objective was to minimize processing costs. Since verifying "outsider" firms (those without prior 7(a) history) entails significantly higher marginal costs than processing legacy borrowers, a flat fee structure implicitly penalizes lenders for serving the most vulnerable firms. Future emergency facilities should adopt a Tiered Origination Fee Structure that compensates lenders based on the complexity of verification rather than just loan size. Lenders should receive a "discovery premium" (e.g., a higher percentage fee) for originating loans to first-time borrowers or underserved demographics. By explicitly pricing the "information friction," the government can align the banks' profit motive with the social goal of equitable distribution, ensuring that the extra effort required to reach non-networked firms is economically viable for intermediaries.

## C. Contribution to the PPP and Small-Business Finance Literature

This study adds to the rapidly growing literature on the design and consequences of PPP by shifting the focus from "where the money went" to "through which institutional channels it flowed."

First, I place prior SBA 7(a) borrowing at the center of the analysis as a key state variable in small-business finance. Whereas most work emphasizes firmographics, geography, banks, or race and community characteristics, I show that participation in a long-standing SBA program strongly conditions access to a new crisis instrument launched years later. Using entropy-balancing weights and propensity-score matching as a robustness check, I find that prior 7(a) borrowing raises the probability of receiving a PPP loan by about 30 percentage points, far larger than the effect of moving a firm several deciles up the credit-risk distribution. This "program-layering premium" is race-neutral in definition but access-unequal in effect, adding an institutional channel to debates on fairness and targeting in PPP allocation.

Second, I recast PPP as a dynamic policy that shifted from relationship-driven triage to more inclusive allocation. Much empirical work evaluates PPP using a single cross-section or focuses on the initial lending waves. By comparing April 2020 with June 2021 in a consistent firm-level framework, I show that the large 7(a) advantage and strong selection on credit quality essentially disappear by mid-2021, on both the extensive margin (who receives a loan) and the intensive margin (loan amounts). This time-variation provides micro-level evidence that reforms, such as priority windows for very small borrowers and relaxed documentation, did more than expand volume; they changed who the marginal dollar reached. The article thus complements macro-level assessments of PPP's aggregate employment and firm-size effects by documenting how its targeting mechanics evolved over the life of the program.

Third, the data and methods offer a portable template for evaluating emergency business credit using observational microdata. I combine near-universe coverage of U.S. firms from Dun & Bradstreet with SBA 7(a) and PPP administrative records, constructing firm-level measures of risk, size, ownership, organizational form, and legacy program participation at an unusually fine scale. Methodologically, I apply entropy-balancing weights to estimate average treatment effects on the treated of prior 7(a) participation and benchmark these against linear-probability and propensity-score-matching estimators. While other PPP studies exploit geographic or bank-level variation and difference-in-differences designs, this article shows how modern reweighting tools can approximate experimental comparisons between firms with and without prior program ties in a purely observational setting. The framework can be adapted to study layering and spillovers in other SBA initiatives, state credit guarantees, or future crisis-lending facilities.

By foregrounding legacy SBA relationships, tracing their time-varying influence, and demonstrating a scalable causal-weighting design, the article advances the PPP literature from documenting ex post distributions toward understanding the institutional channels that grant some firms privileged access to public credit in times of crisis.

## Section VII. Conclusions

This article leverages a near-universe panel of 26 million U.S. firms to analyze how legacy credit relationships conditioned the allocation of emergency liquidity. By linking SBA administrative records with commercial microdata, I document that the "first-come, first-served" design of the Paycheck Protection Program effectively devolved into a relationship-based rationing mechanism during the crisis onset.

Three key insights emerge. First, information frictions dictated the initial allocation. In April 2020, banks utilized prior SBA 7(a) participation as a primary screening technology: firms with legacy ties were approximately 29 percentage points more likely to receive funding than observationally identical non-7(a) firms, a causal estimate derived via entropy balancing. Second, this "insider advantage" was dynamic. As the program matured and policy rules evolved, the predictive power of legacy ties and credit scores dissipated, indicating that the initial rationing was a function of administrative bottlenecks rather than a permanent structural barrier. Third, while the program eventually achieved broad coverage, the reliance on private intermediaries created a "program-layering premium," where participation in peacetime guarantees inadvertently privileged firms for future emergency aid.

These findings offer a clear roadmap for future crisis-response architecture. The trade-off between speed and equity is not inevitable; it is a consequence of relying on commercial underwriting rails for public transfers. To mitigate this, future policy should decouple verification from historical lending relationships. This requires investing in Digital Public Infrastructure for ex-ante verification and establishing pre-qualification mechanisms based on administrative tax data. By reducing the marginal cost of verifying "informationally opaque" firms, policymakers can ensure that the next injection of emergency liquidity reaches the most vulnerable businesses as rapidly as the most connected ones.

# Tables

## Table 1. Summary of Firm Characteristics in April 2020

| | PPP Recipients | | | Non-Recipients | | |
|---|---|---|---|---|---|---|
| Variable | Median | Mean | Std err | Median | Mean | Std err |
| **Independent Variables** | | | | | | |
| 7a Experience | 0.000 | 0.032 | 0.177 | 0.000 | 0.004 | 0.060 |
| Commercial Credit Ranking | 70.000 | 60.844 | 32.198 | 37.000 | 42.277 | 27.847 |
| Financial Stress Ranking | 51.000 | 50.948 | 27.273 | 33.000 | 36.634 | 23.315 |
| Firm size | 4.000 | 10.431 | 78.178 | 2.000 | 5.207 | 68.073 |
| log(firm size) | 1.609 | 1.736 | 0.892 | 1.099 | 1.405 | 0.660 |
| Firm age | 13.000 | 19.341 | 19.120 | 7.000 | 12.210 | 12.782 |
| **Control Variables** | | | | | | |
| Ownership | 0.000 | 0.092 | 0.288 | 0.000 | 0.022 | 0.146 |
| Business Failure History | 0.000 | 0.422 | 0.494 | 0.000 | 0.143 | 0.350 |
| Foreign Ownership | 0.000 | 0.002 | 0.049 | 0.000 | 0.001 | 0.030 |
| Public Status | 0.000 | 0.001 | 0.022 | 0.000 | 0.002 | 0.013 |
| County Population | 9.000 | 7.820 | 1.584 | 9.000 | 7.937 | 1.491 |
| Annual Sales ($) | 165,249 | 5,628,610 | 386,381,894 | 94,459 | 1,636,740 | 241,172,630 |
| log(sales) | 12.015 | 12.373 | 1.749 | 11.456 | 11.384 | 1.389 |
| Subsidiary | 0.000 | 0.009 | 0.092 | 0.000 | 0.008 | 0.090 |
| Metropolitan Location | 0.000 | 0.055 | 0.229 | 0.000 | 0.011 | 0.104 |
| Rural/Suburb Location | 0.000 | 0.040 | 0.195 | 0.000 | 0.009 | 0.092 |
| Industry Location | 0.000 | 0.025 | 0.155 | 0.000 | 0.004 | 0.061 |
| **Sensitivity Test Dependent** | | | | | | |
| Approved Loan Amount ($) | 28,539 | 116,282 | 376,609 | | | |

*Notes:* This table summarizes firm characteristics in April 2020. The dataset includes 26,704,411 firms, of which 22,907,491 received a PPP loan and 3,796,920 did not. These figures provide a descriptive snapshot of PPP participation during the program's first month. Significance levels are denoted by *, **, and ***, corresponding to the 10%, 5%, and 1% levels, respectively.

Table 2. Linear probability model of PPP Recipiency (April 2020)

| Variables | (1) | (2) | (3) | (4) | (5) | (6) |
|---|---|---|---|---|---|---|
| 7a Experience | 0.388*** | | | | | 0.316*** |
| | (0.004) | | | | | (0.006) |
| Commercial Credit Ranking | | 0.002*** | | | | 0.001 * |
| | | (0.000) | | | | (0.000) |
| Financial Stress Ranking | | | 0.002*** | | | 0.001 * |
| | | | (0.000) | | | (0.000) |
| log(firm size) | | | | 0.05 *** | | 0.004** |
| | | | | (0.000) | | (0.002) |
| Firm age | | | | | -0.001*** | 0.019 |
| | | | | | (0.000) | (0.016) |
| Ownership | 0.048*** | 0.035*** | 0.036*** | 0.033*** | 0.055*** | 0.038*** |
| | (0.002) | (0.000) | (0.002) | (0.000) | (0.002) | (0.002) |
| Business Failure History | 0.126*** | 0.103*** | 0.098*** | 0.123*** | 0.137*** | 0.080* |
| | (0.001) | (0.000) | (0.001) | (0.000) | (0.001) | (0.017) |
| Foreign Ownership | 0.098*** | 0.093*** | 0.091*** | 0.101*** | 0.094*** | 0.077*** |
| | (0.011) | (0.002) | (0.011) | (0.002) | (0.011) | (0.029) |
| Public Status | -0.169*** | -0.141*** | -0.146*** | -0.202*** | -0.176*** | -0.102* |
| | (0.023) | (0.004) | (0.023) | (0.004) | (0.023) | (0.062) |
| County Population | 0.001*** | 0.003*** | 0.002*** | 0.001*** | 0.001*** | 0.000 |
| | (0.000) | (0.000) | (0.000) | (0.000) | (0.000) | (0.000) |
| log(sales) | 0.034*** | 0.030*** | 0.032*** | 0.021*** | 0.036*** | 0.029 |
| | (0.000) | (0.000) | (0.000) | (0.000) | (0.000) | (0.020) |
| Subsidiary | -0.210*** | -0.195*** | -0.192*** | -0.247*** | -0.219*** | -0.121*** |
| | (0.004) | (0.001) | (0.004) | (0.001) | (0.004) | (0.003) |
| City Central Region | 0.113*** | 0.117*** | 0.118*** | 0.100*** | 0.115*** | 0.072*** |
| | (0.003) | (0.001) | (0.003) | (0.001) | (0.003) | (0.006) |
| Rural/Suburb Region | 0.085*** | 0.081*** | 0.083*** | 0.066*** | 0.088*** | 0.043 |
| | (0.003) | (0.001) | (0.003) | (0.001) | (0.003) | (0.038) |
| Industry Region | 0.159*** | 0.168*** | 0.173*** | 0.139*** | 0.164*** | 0.045 |
| | (0.004) | (0.001) | (0.004) | (0.001) | (0.004) | (0.048) |
| $R^2$ | 0.107 | 0.112 | 0.108 | 0.103 | 0.098 | 0.216 |

*Notes:* This table reports estimates from model (1a) and (1b) for first-draw PPP loans in April 2020. All columns use 26,704,411 observations. Column (6) includes all firm characteristics (7(a) experience, credit ranking, financial stress ranking, log firm size, and firm age) simultaneously. Columns (1) – (5) report specifications that include only one of these characteristics at a time, always with the same set of controls. All regressions include state and industry fixed effects. Significance levels are denoted by *, **, and ***, corresponding to the 10%, 5%, and 1% levels, respectively.

Table 3. Linear probability model of PPP Recipiency (June 2021)

| Variables | (1) | (2) | (3) | (4) | (5) | (6) |
|---|---|---|---|---|---|---|
| 7a Experience | -0.009*** | | | | | -0.008*** |
| | (0.003) | | | | | (0.003) |
| Commercial Credit Ranking | | 0.000*** | | | | 0.000** |
| | | (0.000) | | | | (0.000) |
| Financial Stress Ranking | | | 0.000*** | | | 0.000* |
| | | | (0.000) | | | (0.000) |
| log(firm size) | | | | -0.001*** | | -0.000 |
| | | | | (0.000) | | (0.000) |
| Firm age | | | | | 0.000 | 0.000 |
| | | | | | (0.000) | (0.000) |
| Ownership | -0.005*** | -0.007*** | -0.007*** | -0.005*** | -0.005*** | -0.002 |
| | (0.002) | (0.002) | (0.002) | (0.002) | (0.002) | (0.002) |
| Business Failure History | 0.011*** | 0.008*** | 0.008*** | 0.011*** | 0.011*** | 0.002* |
| | (0.001) | (0.001) | (0.001) | (0.001) | (0.001) | (0.001) |
| Foreign Ownership | 0.035*** | 0.034*** | 0.034*** | 0.035*** | 0.035*** | 0.028*** |
| | (0.008) | (0.008) | (0.008) | (0.008) | (0.008) | (0.004) |
| Public Status | 0.203*** | 0.204*** | 0.204*** | 0.204*** | 0.203*** | 0.274*** |
| | (0.016) | (0.016) | (0.016) | (0.016) | (0.016) | (0.038) |
| County Population | -0.003*** | -0.002*** | -0.003*** | -0.003*** | -0.003*** | 0.002 |
| | (0.000) | (0.000) | (0.000) | (0.000) | (0.000) | (0.004) |
| log(sales) | 0.000* | 0.000 | 0.000 | 0.000*** | 0.000** | 0.001* |
| | (0.000) | (0.000) | (0.000) | (0.000) | (0.000) | (0.000) |
| Subsidiary | 0.039*** | 0.041*** | 0.041*** | 0.040*** | 0.039*** | 0.026* |
| | (0.003) | (0.003) | (0.003) | (0.003) | (0.003) | (0.016) |
| City Central Region | 0.004** | 0.004** | 0.004** | 0.004** | 0.004** | 0.002** |
| | (0.002) | (0.002) | (0.002) | (0.002) | (0.002) | (0.001) |
| Rural/Suburb Region | -0.006** | -0.006** | -0.006** | -0.005** | -0.005** | 0.005 |
| | (0.002) | (0.002) | (0.002) | (0.002) | (0.002) | (0.004) |
| Industry Region | -0.008** | -0.009*** | -0.008** | -0.008*** | -0.009*** | -0.006 |
| | (0.003) | (0.003) | (0.003) | (0.003) | (0.003) | (0.005) |
| $R^2$ | 0.014 | 0.015 | 0.014 | 0.014 | 0.014 | 0.063 |

Notes: This table reports estimates from model (1a) and (1b) for first-draw PPP loans in June 2021. All columns use 28,172,912 observations. Column (6) includes all firm characteristics (7(a) experience, credit ranking, financial stress ranking, log firm size, and firm age) simultaneously. Columns (1)–(5) report specifications that include only one of these characteristics at a time, always with the same set of controls. All regressions include state and industry fixed effects. Significance levels are denoted by *, **, and ***, corresponding to the 10%, 5%, and 1% levels, respectively.

Table 4. Covariate Balance for Firms With and Without Prior 7(a) Loans (June 2021)

| | Panel A. Raw differences (unweighted) | | | | |
|---|---|---|---|---|---|
| Variable | Mean, 7(a)=1 | Mean, 7(a)=0 | std, 7(a)=1 | std, 7(a)=0 | Std. diff ×100 |
| Commercial Credit Ranking (1–100) | 66.223 | 45.245 | 30.91 | 27.569 | 71.629 |
| Financial Stress Ranking (1–100) | 53.245 | 35.812 | 27.818 | 21.916 | 69.616 |
| log(firm size) | 1.792 | 1.327 | 0.847 | 0.627 | 62.383 |
| Firm age (years) | 21.891 | 12.676 | 18.929 | 12.143 | 57.949 |
| Ownership (=1) | 0.104 | 0.02 | 0.282 | 0.142 | 37.619 |
| Business failure history (=1) | 0.212 | 0.198 | 0.504 | 0.354 | 3.217 |
| Foreign ownership (=1) | 0.002 | 0.001 | 0.051 | 0.032 | 2.344 |
| Public status (=1) | 0.001 | 0.001 | 0.024 | 0.013 | 0 |
| log(annual sales) | 12.532 | 10.412 | 1.697 | 1.361 | 137.838 |
| Metropolitan location (=1) | 0.462 | 0.011 | 0.22 | 0.103 | 262.737 |
| Rural/suburban location (=1) | 0.041 | 0.083 | 0.193 | 0.09 | 27.877 |
| | Panel B. After propensity-score weighting / entropy balancing | | | | |
| Variable | Mean, 7(a)=1 | Weighted mean, 7(a)=0 | std, 7(a)=1 | Weighted std, 7(a)=0 | Std. diff ×100 |
| Commercial Credit Ranking (1–100) | 66.223 | 64.291 | 30.91 | 30.632 | 6.279 |
| Financial Stress Ranking (1–100) | 53.245 | 52.121 | 27.818 | 25.647 | 4.201 |
| log(firm size) | 1.792 | 1.758 | 0.847 | 0.726 | 4.309 |
| Firm age (years) | 21.891 | 20.35 | 18.929 | 17.895 | 8.366 |
| Ownership (=1) | 0.104 | 0.101 | 0.282 | 0.263 | 1.1 |
| Business failure history (=1) | 0.212 | 0.211 | 0.504 | 0.49 | 0.201 |
| Foreign ownership (=1) | 0.002 | 0.002 | 0.051 | 0.042 | 0 |
| Public status (=1) | 0.001 | 0.001 | 0.024 | 0.018 | 0 |
| log(annual sales) | 12.532 | 12.398 | 1.697 | 1.667 | 7.968 |
| Metropolitan location (=1) | 0.462 | 0.454 | 0.22 | 0.187 | 3.918 |
| Rural/suburban location (=1) | 0.041 | 0.04 | 0.193 | 0.129 | 0.609 |

*Notes:* This table reports covariate balance between firms with and without prior SBA 7(a) borrowing in the matched sample ($N_{Treated} = 187,762$; $N_{Control} = 27,985,150$). Treatment ($T_i = 1$) is defined as having at least one 7(a) loan approved between 2010 and 2019. Panel A reports unweighted raw differences. Panel B reports differences after applying entropy-balancing weights to the control group to match the first moments of the covariate vector. Standardized differences are calculated following Austin (2009); absolute values below 10 indicate negligible imbalance. See Section IV.B for methodological details.

Table 5. Estimated Effect of Prior 7(a) Borrowing on PPP Recipiency (June 2021)

| Estimator | ATT on PPP recipient (=1) |
|---|---|
| Raw difference in means | 0.410*** (0.001) |
| OLS (baseline LPM with controls) | 0.316*** (0.006) |
| Propensity-score matching (NN, 1:1) | 0.305*** (0.007) |
| Entropy-balancing weights (EB-ATT) | 0.292*** (0.007) |

*Notes:* The table reports average treatment effects (including model (4)) on the treated (ATT) of prior SBA 7(a) borrowing on PPP recipiency, using the full population by June 2021. $T_i = 1$ is defined as having at least one 7(a) loan approved (2010-2019); $Y_i = 1$ indicates firm received at least one PPP loan. The raw difference in means equals
$$\Pr(Y = 1 \mid T = 1) - \Pr(Y = 1 \mid T = 0),$$
where $\Pr(Y = 1 \mid T = 1) \approx 0.607$ and $\Pr(Y = 1 \mid T = 0) \approx 0.196$ based on the 2×2 distribution of PPP and 7(a) status. LPM (1a) estimated on the full sample with industry and state fixed effects and the full set of controls. PSM uses nearest-neighbor 1:1 matching without replacement on the estimated propensity score $\hat{p}(X_i) = \Pr(T_i = 1 \mid X_i)$, where $X_i$ denotes the same set of pretreatment covariates. Entropy-balancing weights (EB-ATT) reweight non-7(a) firms so that their weighted covariate means match those of 7(a) firms, as documented in Table 4, Panel B. Coefficients are reported as ATT effects on the PPP recipiency probability; robust standard errors are in parentheses. The symbols ***, **, and * denote statistical significance at the 1%, 5%, and 10% levels, respectively.

# Figures

Figure 1. Daily Approved Loans over different stages of PPP.

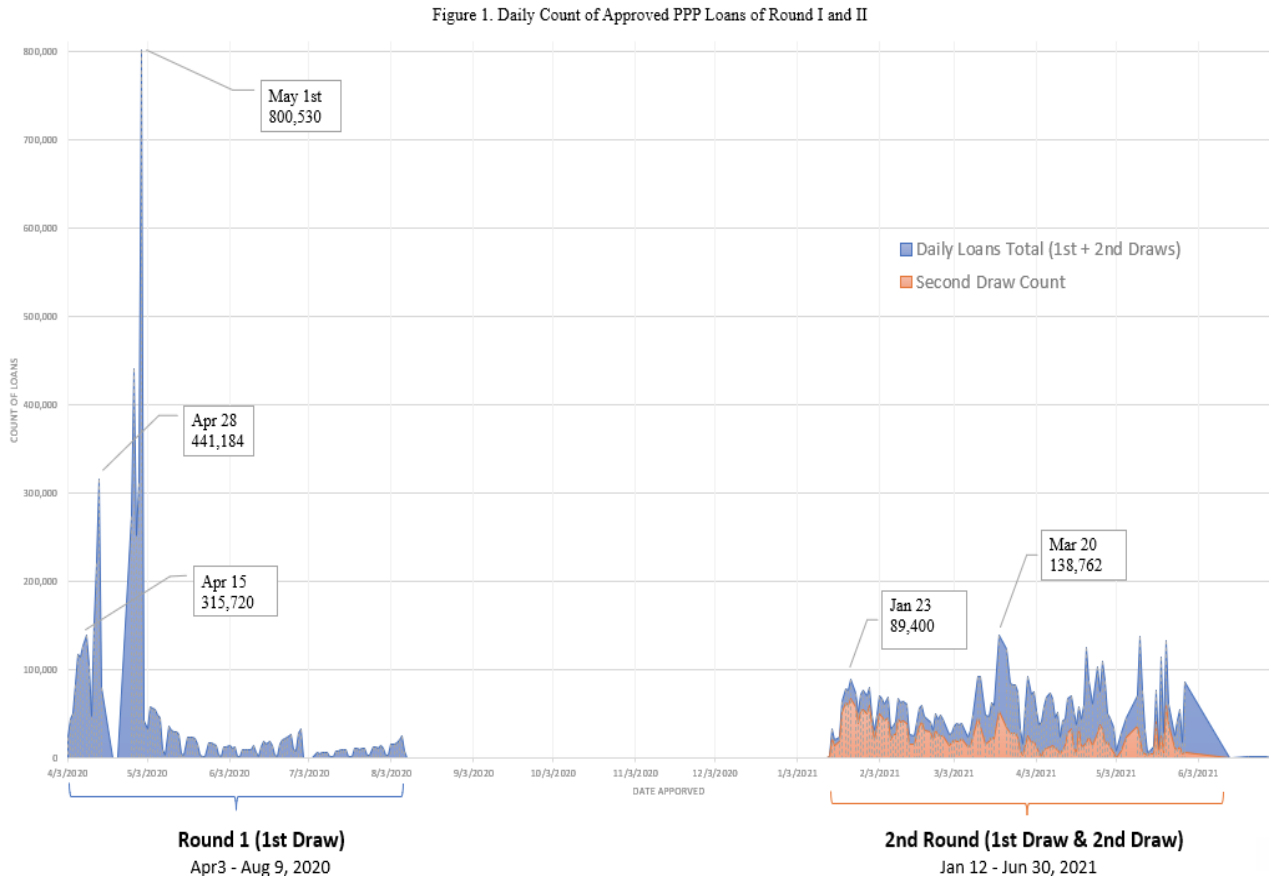

*Notes:* This figure depicts the daily count of approved Paycheck Protection Program (PPP) loans across both major disbursement periods: Round 1 (April–August 2020) and Round 2 (January–June 2021). The blue area represents the total number of daily loan approvals (including both first and second draws), while the orange area identifies second-draw loans only, available exclusively in Round 2. The pattern reveals a highly front-loaded distribution of approvals in each round. During the first round, loan approvals surged immediately after program launch, peaking on May 1, 2020, with 800,530 loans approved in a single day, followed by smaller peaks on April 15 (315,720) and April 28 (441,184). Activity declined sharply after May 2020 as initial funds were exhausted. In the second round, approvals resumed on January 12, 2021, with notable spikes on January 23 (89,400) and March 20 (138,762), driven largely by second-draw loans. The figure underscores the intense initial demand for relief funding and the more gradual, sustained lending pattern during the program's later phase.
Sources: Authors' calculation based on the described data sources.

Figure 2. Binscatters of PPP Loan Recipiency Likelihood vs Business Determinants

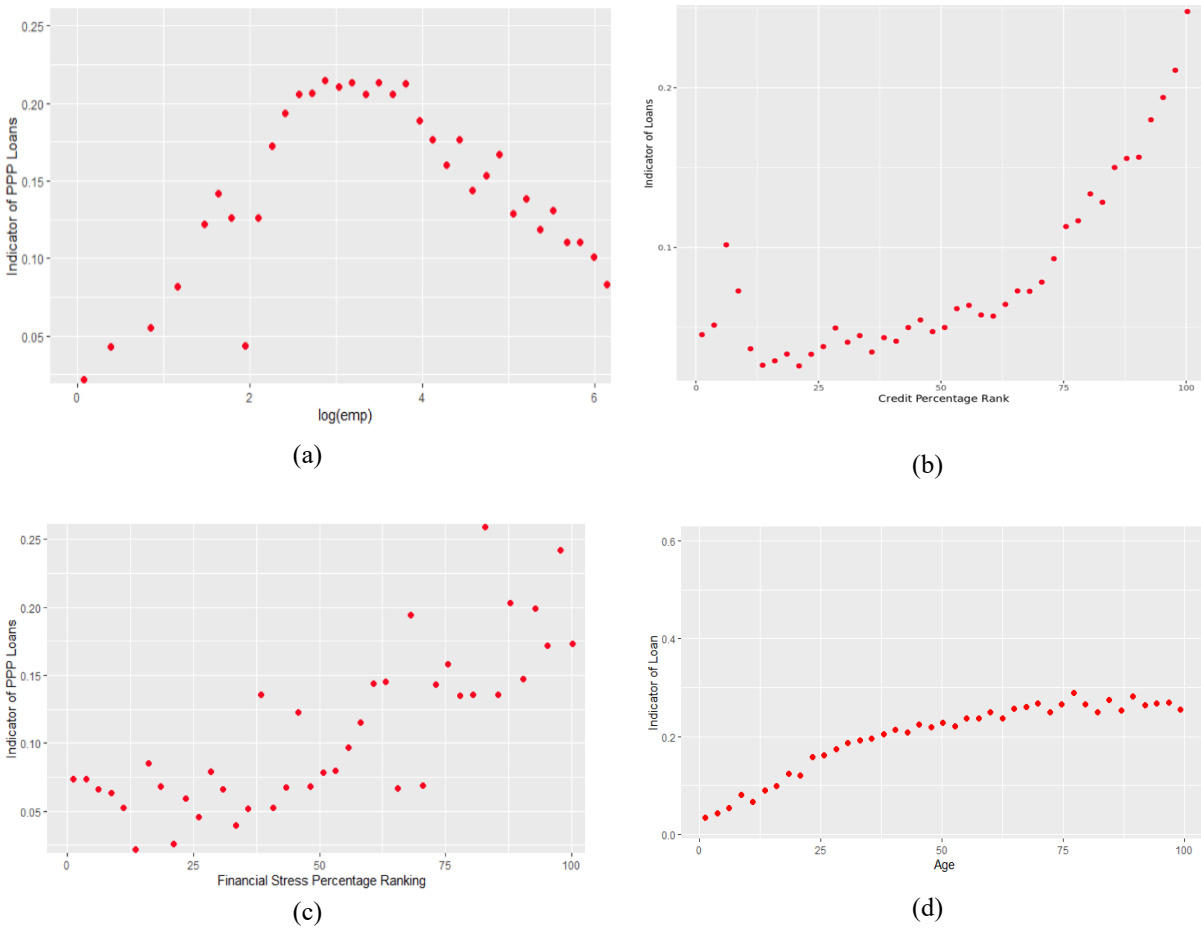

*Notes:* Figure B5 displays four binscatter plots illustrating the relationship between firm or owner characteristics and the probability of receiving a PPP loan. In all panels, the y-axis represents the ratio of firms that successfully obtained a PPP loan within each bin, while the x-axis varies across panels to capture different explanatory dimensions. Panel (a) plots the share of approved firms against the logarithm of firm firm size. The relationship follows an inverted-U pattern, with loan approval rates increasing among small-to-medium firms and gradually declining among the largest firms. Panel (b) shows a strong positive association between firms' credit percentage rank and loan approval probability, indicating that higher-credit firms were substantially more likely to obtain PPP funding. Panel (c) plots approval rates by financial stress ranking. The relationship is relatively flat across most of the distribution but rises in the upper percentiles, suggesting that highly stressed firms exhibited slightly greater loan access, possibly reflecting targeted relief efforts. Panel (d) displays the relationship between firm age and PPP loan approval. The probability of approval rises with firm age up to roughly mid-life firms and then levels off, indicating that older and more established firms had systematically higher access to PPP loans. All panels are constructed from firm-level data, with observations grouped into evenly spaced bins along the x-axis. The plots illustrate broad, unconditional patterns prior to multivariate adjustment and serve as descriptive evidence of heterogeneity in PPP loan accessibility across firm characteristics. Sources: Authors' calculation based on the described data sources.

# Appendix A. Data Construction and Variable Definitions

This appendix documents the construction of the merged PPP–SBA–D&B dataset, the definition and transformation of key variables, and additional descriptive evidence on sample coverage and firm characteristics. It is intended to complement the brief overview in Section III by providing technical details that are not essential for following the main argument.

D&B maintains a large commercial database on U.S. businesses, including firm identifiers, industry codes, employment, location, and a rich set of risk and ownership indicators derived from financial statements, trade-credit histories, and public records. I use D&B snapshots as of April 2020 and June 2021 to represent firms' pre- and mid-program conditions.

PPP loan records, released by the U.S. Small Business Administration, report the name and location of each borrower, loan amount, self-reported jobs retained, and lender identity for all loans approved between April 2020 and mid-August 2021. I merge these records to D&B using standardized firm names, ZIP codes, and industry information to identify which D&B firms received PPP loans and in what amounts.

To measure prior SBA borrowing experience, I use SBA 7(a) loan data from 2010–2019 and match these records to D&B firms. This yields a binary indicator of whether a firm had obtained at least one 7(a) loan prior to the pandemic. The analysis focuses on firms that are active in D&B, have valid identifiers, and possess non-missing values for core risk and employment variables at each reference date. The exact matching rules, exclusion criteria, and sample counts are reported in Appendix A.1–A.4.

## A.1 Construction of SBA 7(a) experience

SBA 7(a) data provide loan-level information on borrower names, locations, loan amounts, and approval dates. I restrict attention to loans approved between 2010 and 2019 to ensure that 7(a) experience predates the COVID-19 shock and PPP rollout. The 7(a) file is matched to D&B firms using the same standardized name and address procedures as for PPP. For each D&B firm, I define:

7(a) experience = 1 if the firm is matched to at least one 7(a) loan during 2010–2019, and 0 otherwise.

In robustness checks (not reported in the main text), I experiment with alternative definitions based on the number of 7(a) loans or cumulative 7(a) volume. These yield similar patterns, underscoring that having any 7(a) history is the most salient dimension for PPP access. Summary statistics on 7(a) participation by industry and state are provided in Appendix Table A2.

## A.2 Risk score ranking

D&B produces proprietary risk measures that estimate the likelihood of severe delinquency, financial distress, or business failure over the coming 12 months. The raw scores are used internally and by lenders but are not directly comparable across time or industries. To make them more interpretable and to facilitate regression analysis, I use them with their percentile rankings:

Commercial Credit Ranking (CCS). Firms are ranked by their Commercial Credit Score within the national cross-section and rescale the ranks to lie between 1 and 100, where 100 corresponds to the lowest predicted probability of severe delinquency and 1 to the highest. This ranking is monotone in the underlying score and can be interpreted as a credit-quality percentile.

Financial Stress Ranking (FS). Similar as CCS, so 100 denotes the lowest predicted probability of failure or severe distress and 1 the highest.

**A.3 Additional descriptive statistics**

Ownership. The ownership binary reflects the ownership status of the firm and is equal to one if the management owns the business, and zero if it rents or leases the business.

Foreign Ownership. The foreign ownership variable is equal to one if the business is owned by a party abroad: it differentiates businesses that are domestically owned and those owned by parent companies based abroad.

Business Failure History. The business failure history variable is a binary equal to one if the firm or its management has had any past bankruptcies or business failures and zero otherwise.

Public and Private. D&B labels each business as public or private and updates this information monthly. Although the PPP is designed for small businesses, some large and publicly listed companies were also eligible for PPP loans.

County Population. The County Population category reflects the residential population of the county where the business is located, providing demographic information on the business's operational environment. This is based on the most recent U.S. Census data, with codes ranging from 0 (denoting populations under 1,000) to 9 (indicating populations of 500,000 and over). A one unit increase in the index indicates a double of population scale.

Subsidiary. The Subsidiary section indicates whether the business is a corporation that is majority-owned by another company.

Location. Location classification provides information about the type of area where the business operates. I categorize areas into four groups: metropolitan, suburban/rural, industrial and other (residential and other types of areas).

Sales. The model utilizes the logarithm of annual sales in USD to control the revenue magnitude.

Industry and State. The model includes 1-digit NAICS industry codes for industry-fixed effects and state fixed effects.

## Appendix B. Robustness Check – Probit Models

In addition to the baseline models (1) and (2), a sensitivity analysis was conducted using a probit model. Probit models, unlike LPMs, do not produce predicted values outside the 0-1 range, making them more suitable for modeling probabilities. I re-estimate equation (1a) using probit (and logit) models for PPP recipiency. For each specification, I report the average marginal effects of $Z_i$ evaluated at the sample means. The signs, magnitudes, and significance levels of the marginal effects are nearly identical to the LPM coefficients, confirming that the results are not driven by the linear probability functional form. The probit model can be expressed as follows

$$P(Y_{ijs} = 1 \mid D_i, X_{ik}, \alpha_j, \rho_s) = \Phi(a + \beta D_i + \sum_k \gamma_k X_{ik} + \alpha_j + \rho_s) \qquad (4)$$

is the population Probit model, where $\Phi$ denotes the cumulative distribution function of the standard normal distribution. It's important to note that the probit model assumes that the error term $\epsilon_i$ follows a standard normal distribution. The model estimates the probability that $Y_{ijs}$ equals 1 given the independent and control variables.

Table B1. Summary of Firm Characteristics in June 2021

|  | PPP Recipients | | | Non-Recipients | | |
|---|---|---|---|---|---|---|
| Variable | Median | Mean | Std err | Median | Mean | Std err |
| **Independent Variables** | | | | | | |
| 7a Experience | 0.000 | 0.006 | 0.079 | 0.000 | 0.007 | 0.084 |
| Commercial Credit Ranking | 49.000 | 50.479 | 31.840 | 41.000 | 46.425 | 29.431 |
| Financial Stress Ranking | 37.000 | 42.065 | 25.582 | 33.000 | 39.323 | 23.795 |
| Firm size | 3.000 | 5.442 | 22.095 | 3.000 | 4.081 | 10.726 |
| log(firm size) | 1.386 | 1.207 | 0.917 | 1.386 | 1.219 | 0.838 |
| Firm age | 8.000 | 13.463 | 15.369 | 8.000 | 12.582 | 13.514 |
| **Control Variables** | | | | | | |
| Ownership | 0.000 | 0.031 | 0.174 | 0.000 | 0.027 | 0.163 |
| Business Failure History | 0.000 | 0.199 | 0.399 | 0.000 | 0.163 | 0.369 |
| Foreign Ownership | 0.000 | 0.002 | 0.049 | 0.000 | 0.001 | 0.030 |
| Public Status | 0.000 | 0.001 | 0.031 | 0.000 | 0.000 | 0.012 |
| County Population | 9.000 | 7.681 | 1.824 | 9.000 | 7.960 | 1.470 |
| Annual Sales ($) | 142,073 | 10,410,737 | 509,045,293 | 150,000 | 1,364,023 | 145,627,829 |
| log(sales) | 11.864 | 12.145 | 1.701 | 11.918 | 12.083 | 1.829 |
| Subsidiary | 0.000 | 0.014 | 0.118 | 0.000 | 0.007 | 0.084 |
| Metropolitan Location | 0.000 | 0.018 | 0.133 | 0.000 | 0.014 | 0.119 |
| Rural/Suburb Location | 0.000 | 0.012 | 0.111 | 0.000 | 0.011 | 0.104 |
| Industry Location | 0.000 | 0.006 | 0.077 | 0.000 | 0.006 | 0.075 |
| **Sensitivity Test Dependent** | | | | | | |
| Approved Loan Amount ($) | 16,665 | 20,703 | 115,137 | | | |

Notes: This table summarizes firm characteristics for first-draw PPP loans in June 2021. The full sample includes 29,765,361 firms, of which 1,592,449 received a PPP loan and 28,172,912 did not. These figures provide a descriptive overview of PPP participation in mid-2021. Significance levels are denoted by *, **, and ***, corresponding to the 10%, 5%, and 1% levels, respectively.
Sources: Authors' calculations based on D&B data and SBA loan records.

## Table B2. Probit Model of PPP Recipiency (April 2020)

| Variables | (1) | (2) | (3) | (4) | (5) | (6) |
|---|---|---|---|---|---|---|
| 7a Experience | 1.187*** | | | | | 0.921*** |
| | (0.006) | | | | | (0.008) |
| Commercial Credit Ranking | | 0.007*** | | | | 0.002*** |
| | | (0.000) | | | | (0.000) |
| Financial Stress Ranking | | | 0.007*** | | | 0.006*** |
| | | | (0.000) | | | (0.000) |
| log (firm size) | | | | 0.118*** | | 0.082** |
| | | | | (0.001) | | (0.05) |
| Firm age | | | | | -0.003*** | -0.001 |
| | | | | | (0.000) | (0.000) |
| Ownership | 0.062*** | 0.025*** | 0.023*** | 0.040*** | 0.100*** | 0.052*** |
| | (0.003) | (0.003) | (0.003) | (0.003) | (0.003) | (0.009) |
| Business Failure History | 0.385*** | 0.308*** | 0.279*** | 0.394*** | 0.433*** | 0.124*** |
| | (0.002) | (0.002) | (0.002) | (0.002) | (0.002) | (0.022) |
| Foreign Ownership | 0.486*** | 0.457*** | 0.452*** | 0.511*** | 0.452*** | 0.169*** |
| | (0.019) | (0.019) | (0.019) | (0.019) | (0.016) | (0.046) |
| Public Status | -1.058*** | -0.889*** | -0.914*** | -1.057*** | -1.072*** | -0.323*** |
| | (0.039) | (0.039) | (0.039) | (0.039) | (0.033) | (0.100) |
| County Population | 0.001** | 0.011*** | 0.004*** | 0.001 | -0.000 | 0.000 |
| | (0.001) | (0.001) | (0.001) | (0.001) | (0.000) | (0.000) |
| log(sales) | 0.203*** | 0.172*** | 0.184*** | 0.159*** | 0.213*** | 0.042 |
| | (0.001) | (0.001) | (0.001) | (0.001) | (0.001) | (0.031) |
| Subsidiary | -1.066*** | -0.96*** | -0.958*** | -1.136*** | -1.124*** | -0.620* |
| | (0.009) | (0.009) | (0.009) | (0.009) | (0.008) | (0.351) |
| City Central Region | 0.224*** | 0.250*** | 0.249*** | 0.213*** | 0.225*** | 0.971* |
| | (0.004) | (0.004) | (0.004) | (0.004) | (0.004) | (0.579) |
| Rural/Suburb Region | 0.144*** | 0.142*** | 0.137*** | 0.121*** | 0.155*** | 0.042 |
| | (0.005) | (0.005) | (0.005) | (0.005) | (0.004) | (0.061) |
| Industry Region | 0.320*** | 0.367*** | 0.376*** | 0.304*** | 0.322*** | 0.111 |
| | (0.007) | (0.007) | (0.007) | (0.007) | (0.006) | (0.189) |

Notes: This table reports estimates from model (4) for first-draw PPP loans in April 2020. All columns use 26,704,411 observations. Column (6) includes all firm characteristics (7(a) experience, credit ranking, financial stress ranking, log firm size, and firm age) simultaneously. Columns (1) – (5) report specifications that include only one of these characteristics at a time, always with the same set of controls. All regressions include state and industry fixed effects. Significance levels are denoted by *, **, and ***, corresponding to the 10%, 5%, and 1% levels, respectively.
Sources: Authors' calculations based on D&B data and SBA loan records.

Table B3. Probit Model of PPP Recipiency (June 2021)

| Variables | (1) | (2) | (3) | (4) | (5) | (6) |
|---|---|---|---|---|---|---|
| 7a Experience | -0.078*** | | | | | -0.072*** |
|  | (0.026) | | | | | (0.026) |
| Commercial Credit Ranking | | 0.002*** | | | | 0.003* |
|  | | (0.000) | | | | (0.002) |
| Financial Stress Ranking | | | 0.001*** | | | 0.001* |
|  | | | (0.000) | | | (0.000) |
| log (firm size) | | | | -0.010*** | | 0.006 |
|  | | | | (0.003) | | (0.004) |
| Firm age | | | | | -0.000 | -0.000 |
|  | | | | | (0.000) | (0.001) |
| Ownership | -0.042*** | -0.056*** | -0.052*** | -0.04*** | -0.035** | -0.034** |
|  | (0.014) | (0.014) | (0.014) | (0.014) | (0.014) | (0.020) |
| Business Failure History | 0.097*** | 0.071*** | 0.073*** | 0.094*** | 0.105*** | 0.125 |
|  | (0.006) | (0.006) | (0.007) | (0.006) | (0.007) | (0.097) |
| Foreign Ownership | 0.192*** | 0.187*** | 0.186*** | 0.194*** | 0.19*** | 0.231 |
|  | (0.056) | (0.056) | (0.056) | (0.056) | (0.056) | (0.180) |
| Public Status | 0.954*** | 0.968*** | 0.966*** | 0.968*** | 0.959*** | 0.422 |
|  | (0.102) | (0.102) | (0.102) | (0.102) | (0.101) | (0.308) |
| County Population | -0.019*** | -0.017*** | -0.018*** | -0.019*** | -0.019*** | -0.02* |
|  | (0.002) | (0.002) | (0.002) | (0.002) | (0.002) | (0.002) |
| log(sales) | 0.002* | -0.000 | 0.001 | 0.004*** | 0.002* | 0.002* |
|  | (0.001) | (0.001) | (0.001) | (0.001) | (0.001) | (0.001) |
| Subsidiary | 0.285*** | 0.304*** | 0.302*** | 0.296*** | 0.283*** | 0.282*** |
|  | (0.022) | (0.022) | (0.022) | (0.022) | (0.022) | (0.053) |
| City Central Region | 0.038** | 0.036** | 0.037** | 0.04** | 0.038** | 0.029* |
|  | (0.017) | (0.017) | (0.017) | (0.017) | (0.017) | (0.010) |
| Rural/Suburb Region | -0.046** | -0.052** | -0.051** | -0.044** | -0.043** | -0.059 |
|  | (0.021) | (0.021) | (0.021) | (0.021) | (0.021) | (0.051) |
| Industry Region | -0.072** | -0.075*** | -0.072** | -0.07** | -0.074*** | -0.055 |
|  | (0.029) | (0.029) | (0.029) | (0.029) | (0.029) | (0.049) |

Notes: This table presents the results for probit model (4), pertaining to the first draw of PPP loans in June 2021. The number of observations for the first 6 regressions is 28,172,912. All regressions include state and industry fixed effects. Significance levels are denoted by *, **, and ***, corresponding to the 10%, 5%, and 1% levels, respectively.
Sources: Authors' calculations based on D&B data and SBA loan records.

Table B4. OLS of the Loan Amount (April 2020)

| Variables | (1) | (2) | (3) | (4) | (5) | (6) |
|---|---|---|---|---|---|---|
| 7a Experience | 34812*** (1634) | | | | | 25023*** (1520) |
| Commercial Credit Ranking | | 98*** (6) | | | | 392*** (77) |
| Financial Stress Ranking | | | 17** (7) | | | 98 (82) |
| log (firm size) | | | | 46272*** (304) | | 2722*** (388) |
| Firm age | | | | | -72*** (14) | -212 (184) |
| Ownership | 26219*** (913) | 25465*** (914) | 26073*** (914) | 14229*** (906) | 27200*** (933) | 3503* (2207) |
| Business Failure History | 14169*** (454) | 12873*** (462) | 14047*** (470) | 9766*** (450) | 15551*** (507) | 1404 (920) |
| Foreign Ownership | 55289*** (4620) | 54892*** (4620) | 55104*** (4621) | 57364*** (4568) | 54836*** (4621) | 7252*** (2926) |
| Public Status | 32 (9877) | 1310 (9878) | -49 (9879) | -26506*** (9768) | -178 (9879) | 898 (6089) |
| County Population | 2178*** (114) | 2302*** (114) | 2199*** (114) | 1772*** (113) | 2144*** (114) | 1038 (795) |
| log(sales) | 16790*** (112) | 16519*** (113) | 16803*** (113) | 5215*** (134) | 16927*** (113) | 4546* (2712) |
| Subsidiary | -35361*** (1719) | -34607*** (1720) | -35566*** (1722) | -62076*** (1708) | -36311*** (1722) | -5291*** (9719) |
| City Central Region | 52327*** (1168) | 52667*** (1168) | 52614*** (1168) | 40180*** (1158) | 52586*** (1168) | 2815 (2169) |
| Rural/Suburb Region | 49016*** (1352) | 48873*** (1352) | 49029*** (1352) | 33895*** (1340) | 49440*** (1354) | 9001 (7349) |
| Industry Region | 125208*** (1822) | 125973*** (1822) | 125840*** (1822) | 105273*** (1806) | 125575***  (1822) | 10501* (6203) |
| $R^2$ | 0.064 | 0.064 | 0.063 | 0.085 | 0.064 | 0.183 |

Notes: This table presents the results for linear model (2a) and (2b), pertaining to the first draw of PPP loans in April 2020. The number of observations for the first 6 regressions is 26,704,411. All regressions include state and industry fixed effects. Significance levels are denoted by *, **, and ***, corresponding to the 10%, 5%, and 1% levels, respectively.
Sources: Authors' calculations based on D&B data and SBA loan records.

Table B5. OLS of the Loan Amount (June 2021)

| Variables | (1) | (2) | (3) | (4) | (5) | (6) |
|---|---|---|---|---|---|---|
| 7a Experience | -139.44 (317.43) | | | | | -50.28 (192.30) |
| Commercial Credit Ranking | | 5.19*** (0.99) | | | | 8.22*** (1.13) |
| Financial Stress Ranking | | | 3.98*** (1.30) | | | 12.98*** (4.30) |
| log(firm size) | | | | 471.05*** (36.29) | | 121.76 (98.39) |
| Firm age | | | | | 20.53*** (2.57) | 39.98 (32.06) |
| Ownership | 1482.25*** (181.42) | 1437.52*** (181.62) | 1450.57*** (181.71) | 1418.94*** (181.47) | 1174.24*** (185.47) | 808.83*** (218.45) |
| Business Failure History | 664.42*** (83.84) | 581.89*** (85.25) | 591.66*** (87.01) | 794.74*** (84.41) | 300.35*** (95.35) | 214.53 (184.20) |
| Foreign Ownership | -1013.22 (903.06) | -1027.29 (903.06) | -1029.41 (903.07) | -1112.26 (903.02) | -934.82 (903.09) | -596.25 (601.91) |
| Public Status | 1429.85 (1944.46) | 1465.27 (1944.44) | 1464.76 (1944.47) | 759.62 (1944.98) | 1280.82 (1944.49) | 1067.52 (1287.87) |
| County Population | 25.94 (21.46) | 30.87 (21.48) | 26.21 (21.46) | 23.811 (21.46) | 39.95 (21.53) | 10.11 (16.45) |
| log(sales) | 194.12*** (15.20) | 188.230*** (15.24) | 192.52*** (15.21) | 113.21*** (16.42) | 190.38*** (15.21) | 164.12* (9.91) |
| Subsidiary | 1948.64*** (332.29) | 2002.87*** (332.42) | 1994.44*** (332.58) | 1486.79*** (334.16) | 2060.52*** (332.54) | 1743.28** (871.00) |
| City Central Region | 1547.00*** (235.18) | 1540.74*** (235.16) | 1545.00*** (235.16) | 1446.16*** (235.27) | 1516.90*** (235.18) | 1504.03*** (235.31) |
| Rural/Suburb Region | 1044.94*** (270.67) | 1025.31*** (270.69) | 1029.84*** (270.71) | 944.73*** (270.75) | 911.25*** (271.18) | 741.76*** (730.86) |
| Industry Region | 617.66* (368.05) | 609.70* (368.02) | 619.28* (368.02) | 475.23 (368.15) | 630.89* (368.01) | 441.09 (465.75) |
| R² | 0.001 | 0.001 | 0.001 | 0.001 | 0.001 | 0.0069 |

Notes: This table presents the results for OLS model (2a) and (2b) for PPP loan amounts in June 2021. The number of observations for the first 6 regressions is 28,172,912. All regressions include state and industry fixed effects. Significance levels are denoted by *, **, and ***, corresponding to the 10%, 5%, and 1% levels, respectively.

# Appendix C. SBA 7(a) loans recipiency by firm characteristics and timeline

Table C1. Pre-PPP SBA 7(a) Participation by Firm Characteristics

| | Total firms | 7a Recipients | 7(a) loan (%) |
|---|---|---|---|
| *Panel A. By employment size* | | | |
| Employment | | | |
| 1–5 | 21,174,425 | 128,293 | 0.61 |
| 6–20 | 6,248,267 | 52,896 | 0.84 |
| 21–50 | 447,567 | 5,324 | 1.19 |
| 51–100 | 182,338 | 978 | 0.54 |
| 101–200 | 82,596 | 278 | 0.34 |
| 201–500 | 37,719 | 106 | 0.28 |
| *Panel B. By firm age* | | | |
| Firm age group (years) | | | |
| 1–5 | 4,425,317 | 8,588 | 0.19 |
| 6–10 | 11,991,351 | 83,731 | 0.70 |
| 11–15 | 4,683,242 | 36,149 | 0.78 |
| 16+ | 7,073,002 | 59,407 | 0.84 |
| *Panel C. By pre-PPP financial-stress percentile* | | | |
| Financial-stress percentile | | | |
| <10 | 1,953,991 | 24,333 | 1.25 |
| 11–30 | 8,989,954 | 30,077 | 0.03 |
| 31–50 | 8,887,936 | 45,619 | 0.53 |
| 51–70 | 4,382,218 | 48,602 | 1.11 |
| 71–90 | 3,104,797 | 32,754 | 1.05 |
| 91–99 | 854,016 | 6,490 | 0.76 |
| *Panel D. By pre-PPP commercial-credit percentile* | | | |
| Commercial credit percentile | | | |
| <10 | 2,522,190 | 21,336 | 0.85 |
| 11–30 | 8,806,435 | 21,068 | 0.24 |
| 31–50 | 5,889,887 | 19,428 | 0.83 |
| 51–70 | 4,054,070 | 23,431 | 0.58 |
| 71–90 | 4,270,697 | 56,611 | 1.33 |
| 91–99 | 2,629,633 | 46,001 | 1.75 |
| *Panel E. By log annual sales* | | | |
| Log(sales) | | | |
| 0–5 | 2,178,441 | 5,389 | 0.25 |
| 5–10 | 471,307 | 1,210 | 0.26 |
| 10–15 | 24,959,967 | 170,715 | 0.68 |
| 15–20 | 556,259 | 10,520 | 1.89 |
| >20 | 6,938 | 41 | 0.59 |

Notes: This table reports pre-PPP participation in SBA 7(a) loans by firm characteristics for the complete PPP period. Results are derived from the D&B–PPP–SBA matched dataset. 7a Recipients count firms with at least one 7(a) approval between 2010 and 2019. 7(a) loan (%) equals $n_{7(a)}/n_{\text{obs}}$, expressed in percent. Financial-stress and commercial-credit groups are based on decile bins of the respective national rankings for the firm at the time when it received the 7(a) loan. The change of the stratified variables over year is ignorable during the period of time, thus making ignorable difference for group totals.
Sources: Authors' calculation based on the described data sources.

Table C2. Monthly distribution of PPP recipients and regarding prior 7(a) experience

| Date | Monthly weight of 7(a) and PPP recipients | Monthly weight of PPP recipients |
|---|---|---|
| Apr-20 | 50.23% | 32.45% |
| May-20 | 14.54% | 16.23% |
| Jun-20 | 2.44% | 3.92% |
| Jul-20 | 0.90% | 1.88% |
| Aug-20 | 0.50% | 0.98% |
| 2020 cumulative | 68.60% | 55.47% |
| Jan-21 | 9.14% | 6.10% |
| Feb-21 | 10.14% | 8.98% |
| Mar-21 | 6.75% | 10.94% |
| Apr-21 | 3.60% | 11.29% |
| May-21 | 1.75% | 7.19% |
| Jun-21 | 0.02% | 0.03% |
| 2021 cumulative | 31.40% | 44.53% |

Notes: This table reports the monthly distribution of PPP loan recipients and of PPP recipients with prior SBA 7(a) borrowing in the matched D&B–PPP–7(a) sample. "Monthly weight of 7(a) and PPP recipients" is the share of all firms that both hold a pre-2020 7(a) loan and receive at least one PPP loan whose first PPP approval falls in the given month, summing to 100 percent over all listed months. "Monthly weight of PPP recipients" is defined analogously as the share of all PPP-recipient firms (regardless of 7(a) status) whose first PPP approval occurs in that month, again summing to 100 percent. "2020 cumulative" and "2021 cumulative" rows report the total shares within each calendar year; minor discrepancies from 100 percent reflect rounding.

## Appendix D. Clarification on 7(a) and PPP

The PPP and the 7(a)-loan program are both instrumental in supporting small businesses, but they serve different purposes and have distinct terms and conditions.

One key difference between the two loans is regarding the credit check requirement. While the SBA's traditional 7(a) loan program usually mandates a credit examination for approval, this requirement is currently suspended for PPP loans.

Specifically, the funds from PPP loans should be directed towards payroll costs, employee benefits, mortgage interest, rent, and utilities for accounts that were in effect prior to February 15, 2020. A unique feature of PPP loans is the provision for loan forgiveness. Borrowers can qualify for loan forgiveness if they allocate the loan towards certain expenses, with a stipulation that at least 60% of the funds must be directed towards payroll costs. PPP loans can reach up to $10 million, with the SBA backing 100% of the loan amount. The maturity of PPP loans is 2 years for loans issued before June 5, and 5 years for loans issued on or after that date. Furthermore, PPP loans have a flat interest rate of 1%.

In contrast, SBA 7(a) loans are intended to facilitate business growth during economically stable times. The funds from these loans can be used for a wide array of purposes, including buying real estate, renovation, new construction, purchasing inventory, investing in machinery or equipment, working capital, refinancing debt, and financing the buyout of a partner. Unlike PPP loans, SBA 7(a) loans do not offer loan forgiveness and require collateral for loans of more than $25,000. The maximum amount for SBA 7(a) loans is $5 million, with the SBA guaranteeing 85% for loans up to $150,000 and 75% for loans greater than $150,000. The loan maturity for SBA 7(a) loans can be up to 5 or 10 years, or even 25 years for real estate financing. The interest rates for SBA 7(a) loans are variable, with a ceiling placed by the SBA on the amount that lenders can charge.